\documentstyle[twocolumn,aps]{revtex}

\begin{document}
\twocolumn[\hsize\textwidth\columnwidth\hsize\csname
@twocolumnfalse\endcsname

\title{Origin of room-temperature ferromagnetism in Mn doped semiconducting CdGeP$_2$}
\author{Priya Mahadevan and Alex Zunger}
\address{National Renewable Energy Laboratory, Golden-80401}
\date{\today}

\maketitle

\begin{abstract}
CdGeP$_2$ chalcopyrites doped with Mn have been recently found to exhibit room temperature ferromagnetism. 
Isovalent substitution of the Cd site 
is expected, however, to create {\it antiferromagnetism}, in analogy with the well-known
CdTe:Mn ($d^5$) case. However, chalcopyrite semiconductors exhibit low-energy intrinsic defects. We show 
theoretically how ferromagnetism results from the interaction of Mn with hole-producing intrinsic defects. 
\end{abstract}

]

\narrowtext

Spintronics \cite{spint} - the use of spin rather than charge in electronics - calls for room-temperature
ferromagnetism in a semiconductor.
While there have been several attempts to dope semiconductors with transition metal atoms \cite{alexrev},
most of them did not show magnetic ordering.
The dopant of choice here \cite{ohno} is Mn$^0$ ($d^5s^2$), being formally Mn$^{2+}$ ($d^5$) when substituting
a divalent cation, and Mn$^{3+}$ ($d^4$) when replacing a trivalent cation. Mn on the divalent
cation site of II-VI
semiconductors is attractive because of the significant solubility of the isovalent 
Mn$^{2+}$, but this produces antiferromagnetism \cite{furdyna}.
The reason is that in a ferromagnetic arrangement
all 3$d$ spin-up orbitals are occupied so there is 
no empty orbital on a neighboring Mn atom
for the electron from one Mn atom to hop onto. 
Therefore an antiferromagnetic arrangement of spins at neighboring
Mn sites is favored. The III-V compounds
doped with Mn would have been suitable candidates for ferromagnetism 
as these systems are known \cite{schneider} to have holes ($d^4 \rightarrow d^5$~+~hole)
when Mn$^{3+}$ substitutes the trivalent cation site. 
However, the ferromagnetic transition temperature (T$_c$) in these compounds is limited by the fact that 
Mn$^{3+}$ has low solubility \cite{ohno}, and 
above a critical concentration ($\sim$ 8\% in GaAs) tends to phase separate.
This was not a limitation while doping the II-VI semiconductors. Recently,
Medvedkin {\it et al.} \cite{sato} came up with the novel idea of 
obtaining both high solubility and high $T_c$ in a semiconductor:
replacing Cd sites in the  chalcopyrite
CdGeP$_2$ with Mn atoms. Although CdGeP$_2$ is isovalent to the
III-V compounds \cite{jaffe} (the average valence of Cd and Ge is a group III element),
a large amount ($\sim$ 50\% at the surface) 
of Mn could be doped into this system, resulting in ferromagnetism in 
a semiconductor at the unprecedented high temperature ($T_c$=320~K) \cite{sato}.
However, the existence of ferromagnetism in CdGeP$_2$:Mn is surprising
since, as explained above, replacement of a divalent site by Mn$^{2+}$ ($d^5$)
is known \cite{furdyna} to give rise to an antiferromagnetic 
interaction between the Mn atoms. Indeed, recent electronic 
structure calculations \cite{freeman} 
found that Mn doped CdGeP$_2$ was antiferromagnetic, just like Mn doped 
CdTe \cite{wei}. We show here that Mn doping in CdGeP$_2$ can produce stable ferromagnetism. 
This hitherto unexplored unusual form of ferromagnetism  results from the interaction of a magnetic ion
such as Mn with holes produced by {\it intrinsic defects}.
The central point is that
chalcopyrite semiconductors are known \cite{zhang} 
to be stabilized by certain intrinsic defects such as cation (Cd,Ge) vacancies, vacancy-antisite pairs
and the presence of hole-producing defects which could result in ferromagnetism being favored
even when Mn dopes the Cd site.  The theoretical 
challenge is to identify the  hole-producing defects that form stable complexes 
with  substitution.  In this 
work we have calculated the formation energies of various kinds of defects and predicted the conditions
favoring the substitution of Cd sites and Ge sites by Mn. 
From this we infer that ferromagnetism is induced by holes generated in structures resulting from the simultaneous
substitution of both Cd and Ge sites or the substitution of only the Ge sites with Mn.
This offers a novel design principle for obtaining both high Mn solubility
and high-$T_c$ ferromagnetism due to defect-induced hole
production in semiconducting systems.

We first calculated the 
formation energies of  intrinsic point defects such as a vacancy at a Cd site (V$_{Cd}$),
a vacancy at a Ge site (V$_{Ge}$) and a Ge antisite on Cd (Ge$_{Cd}$),
finding the energetically stablest defects. Then we calculate formation energies
for Mn substituting a Cd site (Mn$_{Cd}$) and a Ge site (Mn$_{Ge}$).
Finally we deduce which of the many possible complexes among such single 
point defects and Mn would have simultaneously low formation energy
(hence, high concentration) and create hole states (therefore coupling the Mn atoms ferromagnetically).

We performed plane-wave pseudopotential \cite{calc_details,pseudopot,VASP}
calculations with isolated defects/Mn atoms introduced into 64 atom supercells of CdGeP$_2$.
The lattice parameters of these supercells were fixed at the theoretically calculated 
values for CdGeP$_2$: a=5.81~\AA~ and c=10.96~\AA~; while the atomic positions were relaxed.
The formation energy for a defect comprising of atoms $\alpha$ in the charge 
state $q$ was computed using the expression  \cite{zhang}
\begin{eqnarray}
\Delta H_f(\alpha,q)& =& E(\alpha) - E(0) + \sum_{\alpha} n_{\alpha} \mu_{\alpha}^{a} + q (E_{VBM} + \epsilon_f) 
\end{eqnarray}
where  $E(\alpha)$ and $E(0)$ are the total energies of the supercell with and without defect $\alpha$.
$n_{\alpha}$ is the number of each defect atom; 
$n_{\alpha}$~=~-1 if an atom is added, while $n_{\alpha}$~=~1 if an atom is removed.  
$\mu_{\alpha}^a$ is the absolute value of the chemical potential of atom $\alpha$.
Since  the formation energies are conventionally
defined with respect to the elemental solid (s), we express $\mu_\alpha^a$ as the sum 
of a component due to the element in its most commonly occurring 
structure $\mu_{\alpha}^s$, and an excess chemical
potential $\mu_{\alpha}$ {\it i.e} $ \mu_{\alpha}^a$ = $\mu_{\alpha}^s$ + $\mu_{\alpha}$.
Here $\mu_{\alpha}^s$ for P, Ge, Mn and Cd are the total energies 
evaluated for the fully-optimised elemental solids
in the observed crystal structures \cite{pearson}.  
If $\Delta~H_f(CdGeP_2)$ is the formation
energy of CdGeP$_2$, then $\mu_{Cd}$ and $\mu_{Ge}$ are determined by 
\begin{eqnarray}
\mu_{Cd} + \mu_{Ge} + 2\mu_P  & \le &  \Delta~H_f(CdGeP_2) 
\end{eqnarray}
Furthermore,  $\mu_{Cd} \le 0$; $\mu_{Ge} \le 0$, 
because otherwise the elemental solids will precipitate. The presence of other intervening
binary phases, however, further restricts the values of $\mu_{Cd}$ and $\mu_{Ge}$:
One must solve Eq.~(2) alongwith the constraints placed by the formation energies 
$\Delta~H_f(Cd_3P_2)$ and $\Delta~H_f(GeP)$ of Cd$_3$P$_2$ and GeP
\begin{eqnarray}
3\mu_{Cd} + 2\mu_P & \le & \Delta~H_f(Cd_3P_2) \\
\mu_{Ge} + \mu_P & \le & \Delta~H_f(GeP) 
\end{eqnarray}
to find the allowed range for $\mu_{Cd}$ and $\mu_{Ge}$ in CdGeP$_2$.
The electrons ionized upon forming a positively-charged defect join the Fermi sea 
so the formation energy increases by $q~\epsilon_f$, where $\epsilon_f$ is the
fermi energy which varies
from 0~eV at the valence band maximum (VBM) of the
host material to the band gap of the host \cite{calcgap}.
Equations (2)-(4) were solved using  the
experimental values \cite{tables,comment} of the formation energies for the
binary phases Cd$_3$P$_2$ (-1.2~eV) and 
GeP (-0.3~eV), while a value of -1.5~eV, in the same 
range as other chalcopyrites \cite{berger} was used for
CdGeP$_2$. 

The allowed range of chemical potentials $\mu_{Cd}$ and $\mu_{Ge}$
for CdGeP$_2$ and the binaries Cd$_3$P$_2$ and GeP are given in Fig.~1.
There are three distinct chemical potential domains
where CdGeP$_2$ can exist. 
For brevity we represent
each region by one characteristic point:
point A - (Cd rich; Ge poor); point B - (Cd rich; Ge-rich); and point
C - (Cd poor; Ge rich). Figure~2 shows the formation energies of the intrinsic point 
defects $Ge_{Cd}$, $V_{Cd}$ and $V_{Ge}$ as well as substitutional defects Mn$_{Ge}$ and Mn$_{Cd}$
at the chemical potentials A, B and C of Fig.~1 as a function of the Fermi energy. 
The vertical dashed line denotes the GGA gap which is underestimated with respect
to the experimental 1.72~eV gap.
Transition points between charge states  are
indicated by solid circles.
The defects can form acceptor states generating 
holes in the system 
[{\it e.g.} (-/0) or (2-/-)] or donor states that generate electrons [{\it e.g.} (0/+)].
We see that: 
(i)When CdGeP$_2$ is grown at point C (Cd-poor and Ge-rich), and to a lesser extent at point B
(Cd-rich and Ge-rich), the favored substitution of Mn is on the Cd site. This substitution leads 
to a neutral charge state (no holes), thus to AFM.
(ii) However, for $n$-type conditions (E$_F$ near CBM) at point B, and at 
all E$_F$ values at point A (Cd-rich; Ge-poor),
the stablest site for Mn substitution is the Ge site. This substitution forms both single  and
double hole-producing acceptors, which can promote FM. However, under $n$-type conditions at point B,
the charge state q=-2 is favored, which has no holes and therefore promotes AFM. Analysis of the wavefunctions
for different eigenvalues revealed that the Mn 3$d$ levels are located 2-3~eV {\it below} the VBM
and so one spin channel is fully filled (5~electrons). While formally Mn$^{4+}$ would have a $d^3$
configuration, the existence of a fully occupied $d^5$ implies that two electrons are captured from 
the states near $\epsilon_f$. The levels which are emptied correspond to the two acceptor state transitions
taking place at 0.5~eV and 0.57~eV above E$_{VBM}$.
(iii) Cadmium vacancies with the charge state q=-2 are easily formed (in $n$-type conditions)
at points A and C.
(iv) Ge vacancy is stable for $n$-type conditions of point A, though the charge state which has
the lowest energy in these conditions has no holes.
can promote ferromagnetism.
(v) Ge-on-Cd antisite has high formation energy, and would therefore 
not have appreciable concentration.

Having identified the hole-producing centers that can yield ferromagnetism, we next
examine the predicted solubilities of isolated Mn. Our calculated formation 
energies for CdGeP$_2$:Mn and similar calculations for GaAs:Mn show 
consistently lower values (for the appropriate chemical potentials) 
in the former case, predicting higher Mn solubility:
the lowest formation energy of substituting a Ga atom with Mn in GaAs
is 0.5~eV (under Mn-rich, Ga-poor conditions). In contrast, even in the worst-case scenarios,
we find a value of $\Delta~H_f$~0.6~eV for  Mn$_{Cd}$ in CdGeP$_2$
under Cd-rich, Ge-poor conditions. For Mn$_{Ge}$ we find $\Delta~H_f$ $\sim$ 0~eV under Cd-poor, Ge-rich conditions. 

Having established low substitution energies for Mn in CdGeP$_2$ we next see if Mn tends to cluster or not.
We calculated the pairing energy, $\delta$, of a pair relative to infinitely separated
Mn atoms in the host, given, by
$ \delta= [E(2~Mn) - E(0~Mn)] - 2[E(1~Mn) - E(0~Mn)]$
where E($n~Mn$) is the total energy of the supercell with n Mn atoms. A negative
pairing energy would imply that the two Mn 
atoms would prefer to pair rather than stay far apart. The pairing
energy for nearest neighbor  Mn$_{Cd}$ and Mn$_{Ge}$ pairs were found to be
+577~meV and +513~meV respectively for the unit cell considered, implying no clustering. On 
the other hand, Mn substitution of GaAs
had a negative pairing energy, equal to -250~meV for Mn occupying 
nearest neighbor Ga sites.
The corresponding energy for clusters of three or four Mn occupying the vertices of a tetrahedron
surrounding a single As in
GaAs was even more negative, indicating that the Mn atoms showed a tendency
of forming clusters when introduced into GaAs.
The Mn atoms doped into CdGeP$_2$ did not exhibit this tendency.

Having established that the isolated defects Mn$_{Cd}$, Mn$_{Ge}$ and V$_{Cd}$ 
have the lowest formation energy under certain experimental conditions and for some charge states 
the latter two are hole-producing (and therefore 
potentially ferromagnetism promoting), and that Mn$_{Ge}$ pairs repel each other rather than cluster 
we now study various combinations of these point defects in search of stable, hole producing 
{\it complexes}.
Two  Mn atoms were introduced alongwith various intrinsic defects
at various positions of a 16 as well as 32 atom super cells of CdGeP$_2$ corresponding to 
50\% and 25\% Mn respectively. The energies of the ferromagnetic 
and antiferromagnetic arrangements was computed for the fully-optimised super cell ({\it i.e} both
cell external parameters as well as cell-internal atomic positions were relaxed). The formation 
energies of the defect complexes in the 32 atom unit cell 
are plotted in Fig.~3 as a function of $\mu_{Cd}$ in panel (a), and
as a function of $\mu_{Ge}$ in panel (b). The nature of the magnetic ground state (A=AFM and F=FM)
has been indicated in parentheses for the defect complex with the lowest energy. The 
vertical arrows denote the values
of chemical potential for which there is a crossover in the type of defect with the lowest energy. 
We see that:
(i) Under Ge-rich conditions, where {\it isolated}
Mn prefers the Cd site (Fig.~2c), a pair of Mn atoms
also prefers the Cd site (Fig.~3a), leading to an AFM ground state. For 25\% Mn the antiferromagnetic
state is lower in energy than the ferromagnetic state by  105, 19, 13 and 15  meV/Mn for Mn$_{Cd}$-Mn$_{Cd}$
separations of 1st, 2nd, 3rd and 4th neighbors respectively. For the 50\% 
doping, the energy difference changes from 124 to 23~meV/~Mn. The energy difference of 19~meV and
 23~meV for 25\% and 50\% dopings
for the second neighbor
positions is in reasonable agreement with  the difference of 21~meV and 35~meV obtained by earlier 
calculations \cite{freeman}.
(ii) Adding a Cd vacancy to the Mn$_{Cd}$-Mn$_{Cd}$ pair lowers the energy for the Cd-poor 
range of Fig.~3a, but leaves the system as a whole antiferromagnetic.
(iii) Under  Cd-rich conditions, where {\it isolated}
Mn prefers the Ge site (Fig.~2a), a pair of Mn atoms also
prefers the Ge site (Fig.~3b), except at the very Ge-rich end. For 25\% Mn the FM state is lower 
in energy than the AFM state
by 111, 67 and  82 meV/Mn for Mn$_{Ge}$-Mn$_{Ge}$ separations of 1st, 2nd and 3rd neighbors respectively.
For the 50\% Mn case, the ferromagnetic state is still stable by 131 and 158~meV/Mn for 
1st and 2nd neighbors. The fact that the energy difference between the ferromagnetic and the 
antiferromagnetic states is quite large even when Mn atoms are at 3rd neighbor separation 
implies that the magnetic interactions are long-ranged in this case, unlike when Mn dopes 
the Cd site. 
(iv) The combination of Mn$_{Cd}$-Mn$_{Ge}$ pairs is also ferromagnetic, being comparable in energy to when
Mn replaces the Ge sites for certain chemical potentials of Fig.~3b.

In pure CdGeP$_2$ the valence electron configuration is 
$a^2t_p^6$, where $p$ denotes the dominant character of the level. This reflects the fact that Ge contributes 1
electron to each of the 4 bonds with P, Cd contributes ${1}\over{2}$ electron and P contributes ${5}\over{4}$
electrons; the GeP$_4$ tetrahedron transfers one electron to the CdP$_4$ tetrahedron, through the bridging P.
The electronic structure of a transition metal substituting a cation site in a semiconductor 
can be described by the hybridization between the orbitals of the transition metal ion and the anion
dangling bonds formed from the cation vacancy in the host semiconductor \cite{alexrev}. Figure~4 shows this
for Mn$_{Ge}$ in CdGeP$_2$. The levels with $t_{\uparrow}$ 
character on Mn (left) hybridize with the levels with the same symmetry on the Ge-vacancy
dangling bond (right), thus forming a bonding "crystal field resonance" (CFR) 
$t_{CFR\uparrow}$ located deep in the valence band with dominant Mn
character, and antibonding "dangling bond hybrid" (DBH) 
$t_{DBH\uparrow}$ with dominantly $p$ character. A similar hybridization of the
down-spin levels results in the down-spin counterparts of the CFR and DBH levels. 
The CFR level with $e$ symmetry exists too.
These $e_{CFR}$, $t_{CFR}$ levels can hold 5 electrons.
When Mn is substituted on the Ge site, 4 of its 7 valence electrons replace those of Ge, leaving
3 electrons to occupy the $t_{CFR\uparrow}~~e_{CFR\uparrow}$. However, since $t_{DBH}$ is higher 
in energy than $e_{CFR\uparrow}$, two electrons move into $e_{CFR\uparrow}$, leading to 
($e^2_{CFR\uparrow}t^3_{CFR\uparrow}$)
consequently, two holes (empty circles in Fig.~4) 
are generated in the DBH levels.
The intraatomic exchange splitting on the Mn atoms is large, $\sim$ 0.6-0.7~eV \cite{spinpol}, and what is
unusual here is that it induces a large negative (0.8-1.0~eV) exchange splitting on the DBH levels which have dominantly P
$p$ character. The direction of the exchange splitting on the DBH (spin-down below spin-up) 
is opposite to the direction of exchange
splitting on the CFR levels,
and this can be understood by simple tight-binding arguments. If the up-spin states on both the DBH
and CFR are occupied, then the energy gain as a result of hybridization is small and comes from the
single electron occupying the down spin DBH levels. However, if the up spin states on the CFR and the down
spin states on the DBH are occupied, the energy gain as a result of hybridization is huge and there
are contributions from both up and down spin channels. Hence the configuration in which the up spin levels
on the CFR and the down spin levels on the DBH are occupied is the ground state of the system. 
A similar hybridization-induced antiferromagnetic coupling was earlier proposed
to explain the high Curie temperature observed in another ferromagnet in which the magnetic
Fe atoms are separated by nonmagnetic Mo atoms \cite{sfmo}.
In the inset of Fig.~4  we show the contributions to the total DOS from the up-spin states as well as the Mn $d$
states in a small energy window near the fermi energy. The states are found to be strongly spin-polarized,
with just 25\% Mn $d$ character from the two Mn atoms present in the 32 atom super cell considered here.
As a result of the large exchange splitting induced on the DBH states, the downspin DBH bands are 
energetically degenerate with the dominantly P $p$ bands. The strongly polarised as well as delocalised 
down spin DBH band we believe is responsible for the long-ranged magnetic interactions, and, consequently
the room temperature ferromagnetism found in Mn doped CdGeP$_2$.

We thank D.D. Sarma for useful discussions. This work was supported by the U.S. DOE under contract no. DE-AC36-99-G010337.

\begin{figure}
\caption{ The range  of Cd and Ge chemical potentials where CdGeP$_2$, GeP and Cd$_3$P$_2$ are stable.
}
\end{figure}
\begin{figure}
\caption{ The formation energies of different charge states of isolated defects calculated as a 
function of fermi energy in a 64 atom cell of CdGeP$_2$ for three chemical potentials: A, B and C.
The vertical dashed line denotes the calculated GGA band gap. Solid circles denote transition energies
between charged states.
}
\end{figure}
\begin{figure}
\caption{ The formation energies as a function of (a) Cd (b) Ge chemical potentials calculated for
complex defects in a 32 atom cell of CdGeP$_2$. The magnetic ground state (A=AFM, F=FM) for the lowest energy state
is indicated in parentheses. The vertical arrows denote the value of chemical potential for which there
is a crossover in the lowest energy defect type.}
\end{figure}
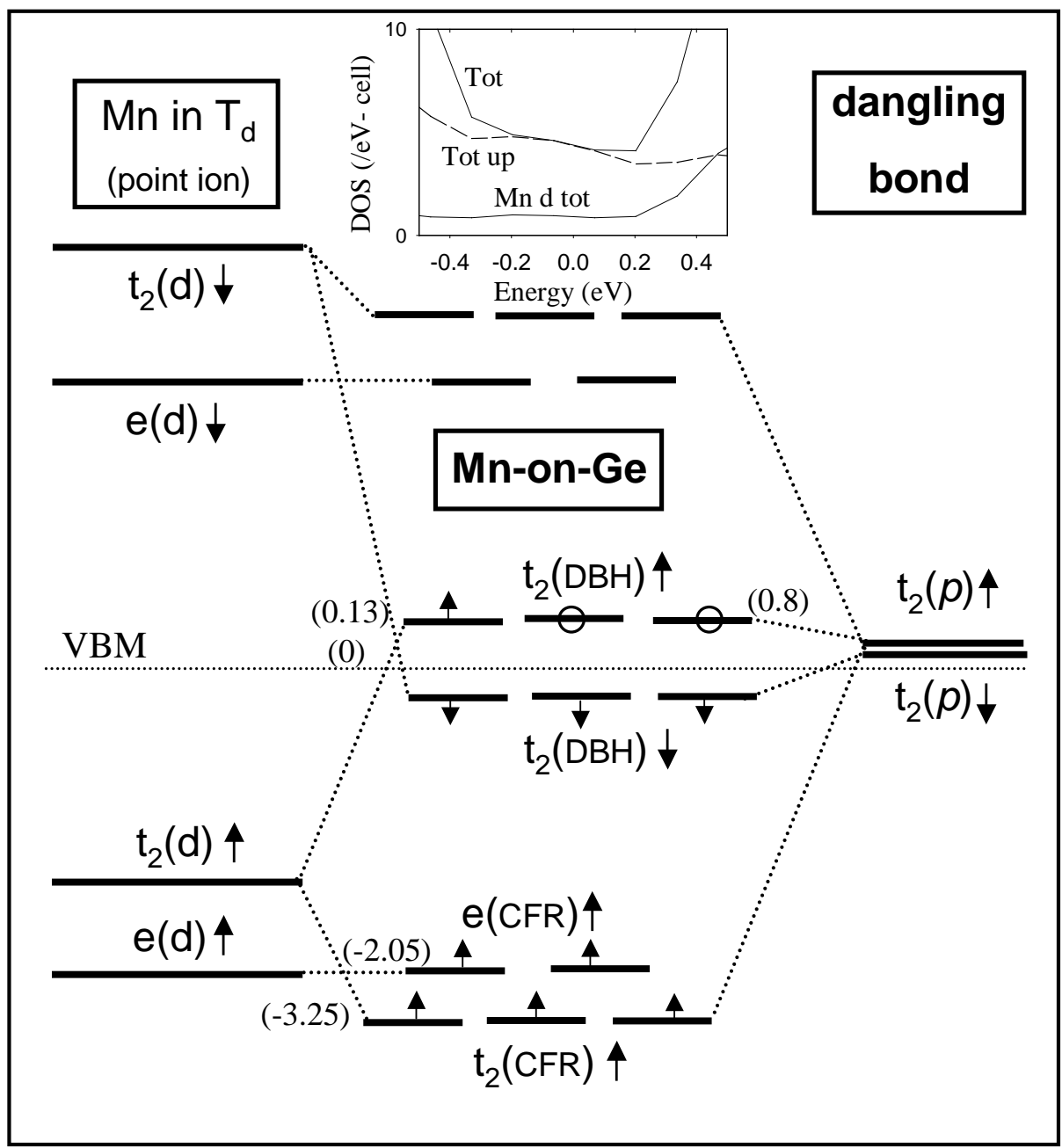
\begin{figure}
\caption{The energy level diagram of Mn$_{Ge}$ in CdGeP$_2$. The energy of the levels are indicated
in brackets. The inset shows the contributions to
the total DOS in the energy interval -0.5~eV to 0.5~eV from the up spin states as well as Mn 3$d$ 
calculated for 2 Mn$_{Ge}$ in a 32 atom unit cell of CdGeP$_2$.
}
\end{figure}

\begin{references}
\bibitem{spint}
T.~Dietl {\it et al.}, Science {\bf 287}, 1019 (2000);
T.~Dietl, J. Appl. Phys. {\bf 89}, 7437 (2001).
\bibitem{alexrev}
A.~Zunger in {\it Solid State Physics} {\bf 39}, 275 (Academic Press, New York, 1986).
\bibitem{ohno}
H.~Munekata {\it et al.}, Phys. Rev. Lett. {\bf 63}, 1849 (1989).
\bibitem{furdyna}
J.K.~Furdyna and J.~Kossut, {\it Semiconductors and semimetals} {\bf 25}, (Academic Press, San Diego, 1988).
\bibitem{schneider}
J.~Schneider {\it et al.}, Phys. Rev. Lett. {\bf 59}, 240 (1987).
\bibitem{sato}
G.A.~Medvedkin {\it et al.}, Jpn. J. Appl. Phys., Part 2 {\bf 39}, L949 (2000).
\bibitem{jaffe}
J.E.~Jaffe and A.~Zunger, Phys. Rev. B {\bf 30}, 741 (1984).
\bibitem{wei}
S.H.~Wei and A.~Zunger, Phys. Rev. B {\bf 35}, 2340 (1987).
\bibitem{freeman}
Yu-Jun Zhao, W.T.~Geng, A.J.~Freeman and T.~Oguchi, Phys. Rev B {\bf 63}, 201202(R) (2001).
\bibitem{zhang}
S.B.~Zhang, S.H.~Wei and A.~Zunger, Phys. Rev. Lett. {\bf 78}, 4059 (1997);
S.B.~Zhang, S.H.~Wei, A.~Zunger and H.~Katayama-Yoshida, Phys. Rev. B {\bf 57}, 9642 (1998).
\bibitem{calc_details}
The pseudopotential plane-wave method \cite{pseudopot} using GGA PW91 
ultrasoft pseudopotentials as implemented in VASP \cite{VASP} was used for the calculations. Cutoff energies
of 16.7~Ry (28.2~Ry) and 15.9~Ry (20.2~Ry) were used for the plane-wave (augmentation charge) for the 
calculations with and without Mn respectively. A Monkhorst-Pack k-points grid of 4x4x2, 2x4x2 and 2x2x2 was used for
the calculations involving 16, 32 and 64 atoms.
\bibitem{pseudopot}
J.~Ihm, A.~Zunger and M.L.~Cohen, J. Phys. C:{\bf 12}, 4409 (1979).
\bibitem{VASP}
G.~Kresse and J.~Furthm$\ddot{u}$ller, Phys. Rev. B. {\bf 54}, 11169 (1996);
G.~Kresse and J.~Furthm$\ddot{u}$ller, Comput. Mat. Sci. {\bf 6}, 15 (1996).
\bibitem{pearson}
P.~Villars and J.L.~Calvert, {\it Pearson's Handbook of Crystallographic Data for Intermetallic Phases}
(ASM International, Materials Park, 1991).
\bibitem{calcgap}
The GGA calculated gap was found to be 0.65~eV.
\bibitem{tables}
Editors Wagman {\it et al.}, {\it Vol. 11, NBS tables of chemical thermodynamic 
properties}, (ACS and AIP for the National Bureau of Standards, 1982).
\bibitem{comment}
The calculated formation energies were -1.3~eV, +0.33~eV and -0.93~eV for Cd$_3$P$_2$, GeP and CdGeP$_2$.
The use of the calculated formation energies instead of the experimental ones does not change the 
conclusions of our work.
\bibitem{berger}
L.I.~Berger, {\it Semiconductor Materials} (CRC press, New York, 1997), p.230.
\bibitem{spinpol}
P.~Mahadevan, N.~Shanthi and D.D.~Sarma, J.~Phys.~Condens.~Matter {\bf 9}, 3129 (1997).
\bibitem{sfmo}
D.D.~Sarma {\it et al.}, Phys. Rev. Lett. {\bf 85}, 2549 (2000).
\end{references}
\end{document}